\documentclass[11pt,a4paper]{article}

\usepackage{hyperref}
\usepackage{graphicx}
\usepackage{courier}
\usepackage{scalefnt}
\usepackage{flushend}
\usepackage{authblk}
\usepackage{algorithm}
\usepackage[noend]{algpseudocode}
\usepackage{amssymb}
\usepackage{fancyhdr}
\usepackage{geometry}
\usepackage{epstopdf}
\usepackage{fancyhdr}

\begin{document}
\title{Quantum Telescope: \ feasibility and constrains}

\author[1,*]{A. R. Kurek}
\author[2]{T. Pieta}
\author[2]{T. Stebel}
\author[1,3]{A. Pollo}
\author[4]{A. Popowicz}

\affil[1]{Astronomical Observatory of the Jagiellonian University, Orla 171, 30-244 Cracow, Poland}
\affil[2]{Faculty of Physics, Astronomy and Applied Computer Science, Jagiellonian University, S. \L{}ojasiewicza 11, 30-348 Cracow, Poland}
\affil[3]{National Centre for  Nuclear Research, Ho\.za 69, 00-681 Warsaw, Poland}
\affil[4]{ Silesian University of Technology, Institute of Automatic Control, Akademicka 16, 44-100 Gliwice, Poland}
\affil[*]{Corresponding author: aleksander.kurek@uj.edu.pl}

\maketitle

\begin{abstract}
\noindent
Quantum Telescope is a recent idea aimed at beating the diffraction limit of spaceborne telescopes and possibly also other distant target imaging systems. There is no agreement yet on the best setup of such devices, but some configurations have been already proposed.
\vspace{1mm}
\\
\noindent
In this Letter we characterize the predicted performance of Quantum Telescopes and their possible limitations. Our extensive simulations confirm that the presented model of such instruments is feasible and the device can provide considerable gains in the angular resolution of imaging in the UV, optical and infrared bands. We argue that it is generally possible to construct and manufacture such instruments using the latest or soon to be available technology. We refer to the latest literature to discuss the feasibility of the proposed QT system design.
\end{abstract}

{\scriptsize Copyright:  \ The Optical Society (OSA Publishing) 2016. Published in Optics Letters Vol. 41 No. 6 (2016)\\}

\noindent
\section{Introduction}
Astronomical images obtained form the ground suffer from serious degradation of resolution, because the light passes through a turbulent medium (the atmosphere)~\cite{Tikhomirov1991} before reaching the detector. A number of methods was developed to alleviate this phenomenon~\cite{AOrev, OTCCD, ImprovResReview, SAA} and multiple special-case solutions were implemented as well (e.g.~\cite{HSTpluto}), but none of them is able to provide a perfect correction and restore the diffraction-limited (DL) image. Therefore, the best observatories, in terms of angular resolution, are the spaceborne ones, since they are limited only by the DL. Recently, there is an increasing interest in the attempts to overcome the DL boundary. Devices which can give such possibility are called the Quantum Telescopes (QT). First QTs will probably work in the UV, optical and IR bands, mainly because of the speed, maturity and reliability of the detectors. Latest progress in Adaptive Optics (AO), especially so called Extreme AO, makes the use of QTs realistic also in ground-based observatories~\cite{ExAO}.

\begin{figure}[t]
	\includegraphics[width=\linewidth, trim=0.25cm 0 0.1cm 0]{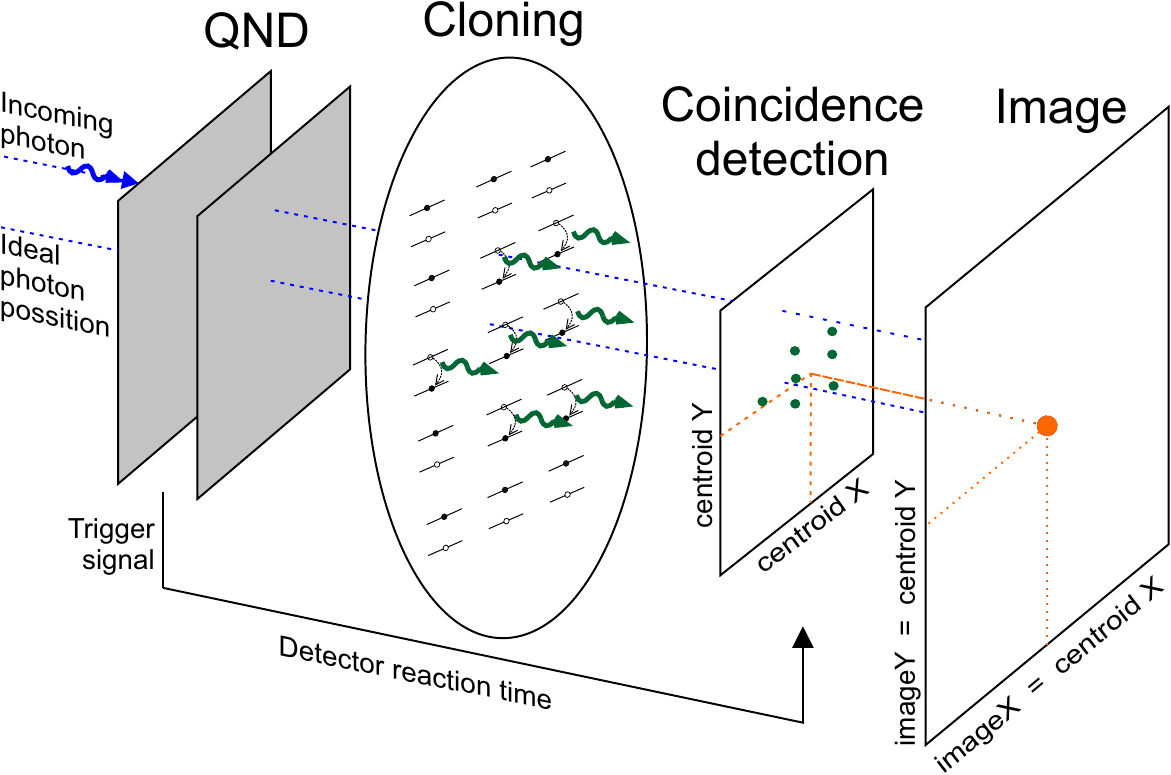}
	\caption{
	Our proposed upgraded design of the Quantum Telescope scheme based on \cite{QT1}. The incoming photon is detected (by the cavity-free QND, if possible \cite{QNDnoCavity, cavityReview}) and a triggering signal turns on a coincidence detector for a short time -- just enough to register the clones of a given photon but not the spontaneous emission. Centroid position of these clones is passed as a pixel value increment [1 ADU] to the image. ``Ideal photon position'' would be obtained by a telescope of an infinite mirror size.
	}
	\label{fig:QTscheme}
\end{figure}

\begin{figure*}
	\includegraphics[width=\linewidth, trim=0.1cm 0.3cm 0 0, bb=0 10 1115 181]{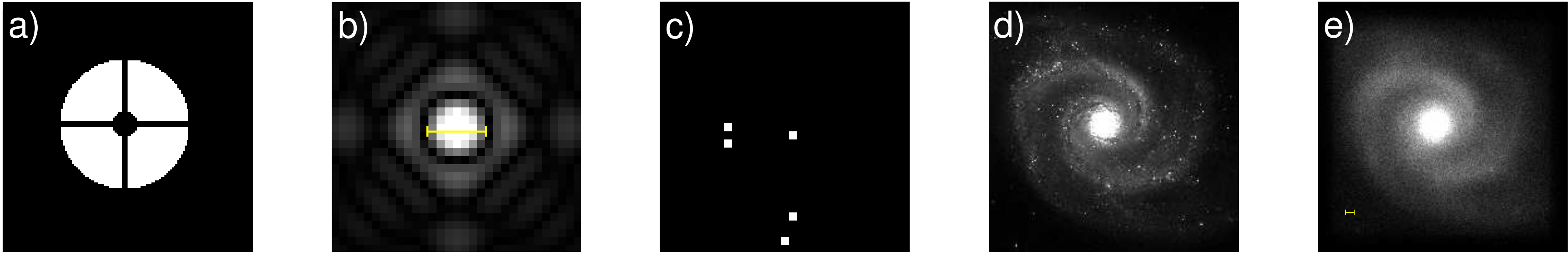}   

	\caption{
	a) obscured pupil used for simulations; \  \ b) decimated diffraction pattern of the pupil; \ \ c) exemplary counts from the diffraction pattern (very short exposure);  \  \ d) input image of a galaxy used for simulations~\cite{M51};\  \ e) CT outcome (diffraction pattern size to the first zero $\lambda$/D = 9 pixels, yellow marker).
	}
	\label{fig:obscuredPupil}
\end{figure*}

\noindent
In this Letter the general idea of Quantum Telescopes is considered. In particular we refer to the setup proposed by~\cite{QT1}, since, to our knowledge, it is the only existing detailed description of a QT. In Fig.~\ref{fig:QTscheme} we propose an upgraded version of this setup. According to~\cite{QT1}, each photon coming from the extended source triggers a signal by QND\footnote{Quantum Non-Demolition (QND) -- a type of measurement of a quantum system in which its state is negligibly changed. See~\cite{QNDrev} for further details.} detection and gets cloned~\cite{cloningRev, cloning1photon, AmplificationRev, Amplification}. The coincidence detector controlled by the trigger is turned on for a short period~\cite{FundLimitsOfDetectors} and registers the clones. After that it is quickly turned off, so that it receives only a small fraction of spontaneous emission from the cloning medium and virtually no clones from other photons. If the source is too bright and emits too many photons per unit of time, a gray filter should be installed. As a result, a set of clones is produced and registered. The centroid position of the clones cloud is used to add 1 ADU\footnote{Analog-Digital Unit is a measure of signal (photoelectrons) in a CCD/CMOS pixel. For simplicity, in our simulations we assumed the unity gain: 1ADU/e$^-$.} at the corresponding position of the high resolution output image. The exposure time has to be much longer than in classic telescopes (CT), since (a) in most cases a narrowband filter has to be applied and (b) the QND detection efficiency is much below 100\%~\cite{Grangier1998}.

\vspace{5mm}
\section{Simulations}
\label{sec:simulations}
\noindent
Below we present the results of our detailed simulations of the QT system, discuss the feasibility of building such a device and predict its expected performance. To our knowledge, this is the first paper describing the detailed simulations of a QT of any design, as this is a preliminary concept.

\vspace{4mm}
\noindent
In our simulations as input images we used parts of real images obtained by the Hubble Space Telescope (HST). Such images are optimal for the purpose of QT testing because this observatory is working at its DL. We cropped and decimated them to 200$\times$200 pixels to speed up the computations. For the comparison, we also simulated the process of digital image formation in the case of a CT (for a review on high angular resolution imaging see~\cite{Saha}) using the same images (Fig.~\ref{fig:obscuredPupil}e). We assumed a real telescope for which the pupil is obscured by a secondary mirror and its truss (``spider``)~\cite{diffPatterns}. We assumed a similar size of a secondary mirror and truss as it is installed on the HST~(Fig.~\ref{fig:obscuredPupil}a).

\begin{figure}[h]
	\centering
	\includegraphics[width=\linewidth, trim=0 0.4cm 0 0.4cm]{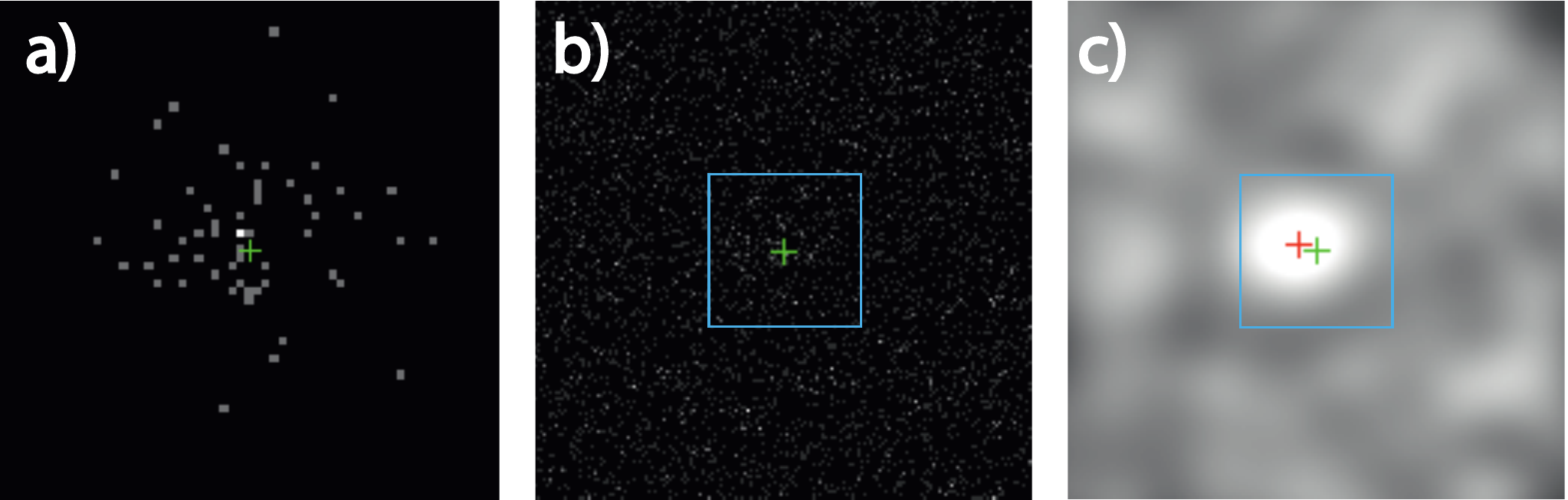}
	\caption{
	Three consecutive steps of the simulations: \ a) clones frame (60 clones in this example), \ b) clones and noise frame together (S/N = 1/7),\ \ c) effect of matched filtration. Green \lq\lq{+}\rq\rq{} denotes real center of the frame, red \lq\lq{+}\rq\rq{} denotes the centroid position obtained from the filtering. The zoomed region presented in the image a) is indicated by blue rectangle in b) and c). With this exemplar parameters, the mean centroid recovery error is 3.8 pixels.
	}
	\label{fig:drawing}
\end{figure}

\vspace{1mm}
\noindent
In the case of a CT, we sampled the counts (photons) from the original image~(Fig.~\ref{fig:obscuredPupil}d) and distributed them according to the Airy diffraction pattern decimated to 61x61 pixels~(Fig.~\ref{fig:obscuredPupil}b). For the simulation of a QT, each photon from the reference image was converted to $N$ cloned photons (Fig.~\ref{fig:obscuredPupil}c). For $N$ we assumed the Poissonian distribution. The clones were arranged using the Gaussian profile (see~\cite{cloning1photon} Fig.~2 therein for justification; we assumed $\sigma$ = 10) centered around the photon's \lq\lq{ideal}\rq\rq position. To include the effects of spontaneous emission, on the top of it we added counts distributed equally  within the coincidence detector plane and governed by the Poissonian distribution. The mean noise level was tuned to achieve a given signal-to-noise ratio (SNR; see exemplary simulated exposure in~Fig. \ref{fig:drawing}b). The assessment of the SNR was based on the comparison of the number of cloned photons with the number of counts originating from the spontaneous emission within the circular aperture of 3$\sigma$ radius (i.e. within the aperture, for which nearly all the clones are received). It follows the approach of SNR derivation presented in~\cite{errata}.

\noindent
In the next stage of simulations of the QT image formation, we computed the centroid of clones employing the Matched Filtering approach\footnote{The MF is an efficient detector of the known template in noisy environments and is widely used in radars or sonars, where the weak, well-defined reflected signals have to be detected~\cite{MF, MF2}.}. As the cloned photons exhibit Gaussian spatial distribution~\cite{cloning1photon}, in our calculations the image registered on the coincidence detector was first convolved with the Gaussian ($\sigma$  = 10) and then the centroid was obtained from the position of the maximum value of such a filtered image~(Fig. \ref{fig:drawing}c).

\begin{algorithm}
\caption{\ \ Simplified QT simulation steps}\label{alg:QTalg}
\begin{algorithmic}[1]
\For{$i \gets 1 \textrm{ to } rows$}
	\For{$j \gets 1 \textrm{ to } columns$}
		\State $C$ = $ADU( \ inputImage$($i$,$j$) \ )
		\For{$k \gets 1 \textrm{ to } N$}
			\State generate $noiseFrame$
			\State generate $N$ clones in the $noiseFrame$ around ($i$,$j$)
			\State compute $centroid$
			\State paste $centroid$ to $hiResIm$($i$,$j$)
		\EndFor
	\EndFor
\EndFor
\end{algorithmic}
\end{algorithm}

\noindent
We ran the simulations for different numbers of clones, reaching also very high numbers (up to the expected value of \textasciitilde10k, see~\cite{DeMartini2015} for justification). The mean level of the Poissonian noise of spontaneous emission was set so that SNR was: 3/1, 2/1, 1, 1/2, 1/3, 1/4, 1/5, 1/6, 1/7, 1/8, 1/9, 1/10, 1/11, 1/12, 1/13 and 1/14. Such a selection of SNR includesa value of  1/7.3 which was assessed for QT in~\cite{errata}.

\noindent
The quality of the simulated QT outcomes was assessed by two indicators: Peak Signal-to-Noise Ratio (PSNR) and mean centroid error (RMS value). For 16-bit pixel representation the PSNR measure is defined as follows:

\begin{equation}
\textrm{PSNR} = 10\cdot\textrm{log}_{10}\bigg(\frac{2^{16}}{\sqrt{MSE}}\bigg)
\end{equation}

\begin{equation}
MSE = \frac{1}{mn}\sum_{i=1}^{m}\sum_{j=1}^{n}\big(I_{QT}(i,j)-I_{ref}(i,j)\big)^2,
\end{equation}

\begin{figure*}
	\centering
	\includegraphics[width=\linewidth, trim=0 1.1cm 0.4cm 0]{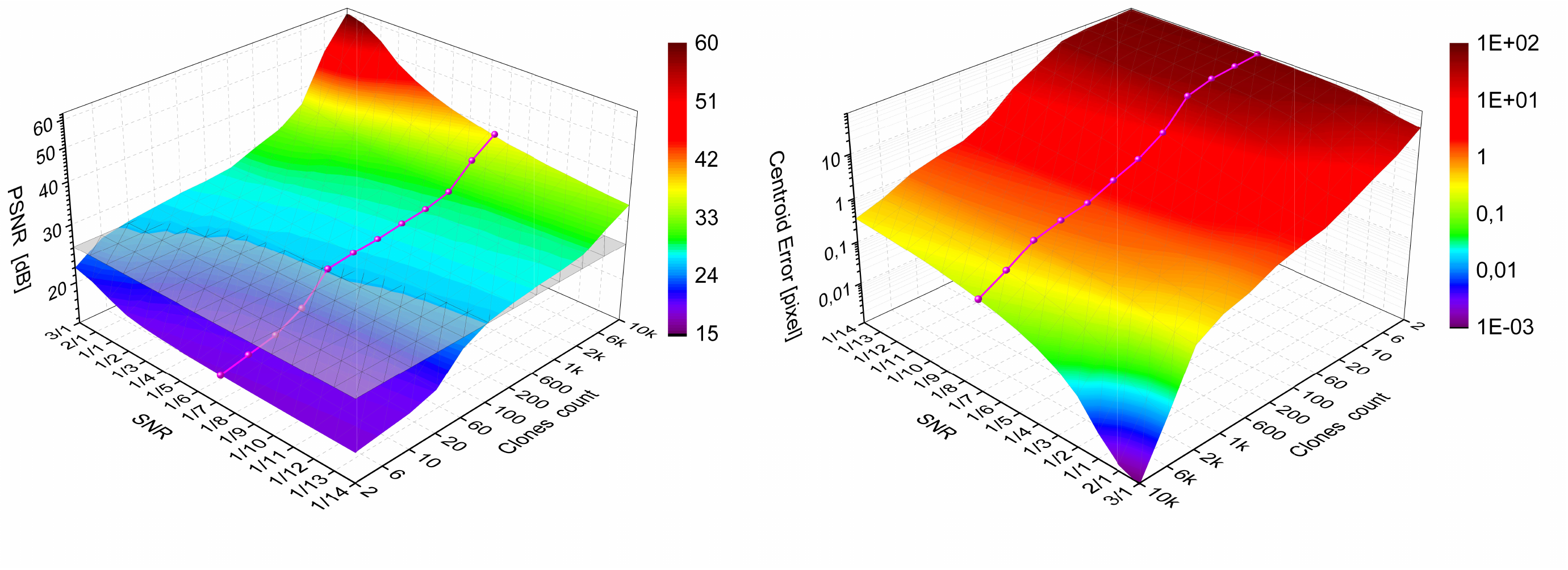}
	\caption{
	QT efficiency diagnostics. Left: PSNR results, violet line represents efficiency for SNR = 1/7.3. Translucent surface in PSNR plot represents CT efficiency. Right: Centroid Error estimation in pixels.
	}
	\label{fig:surfs}
\end{figure*}

\begin{figure}[!ht]
	\centering
	\includegraphics[width=\linewidth, trim=0.3cm 0.5cm 0 0cm]{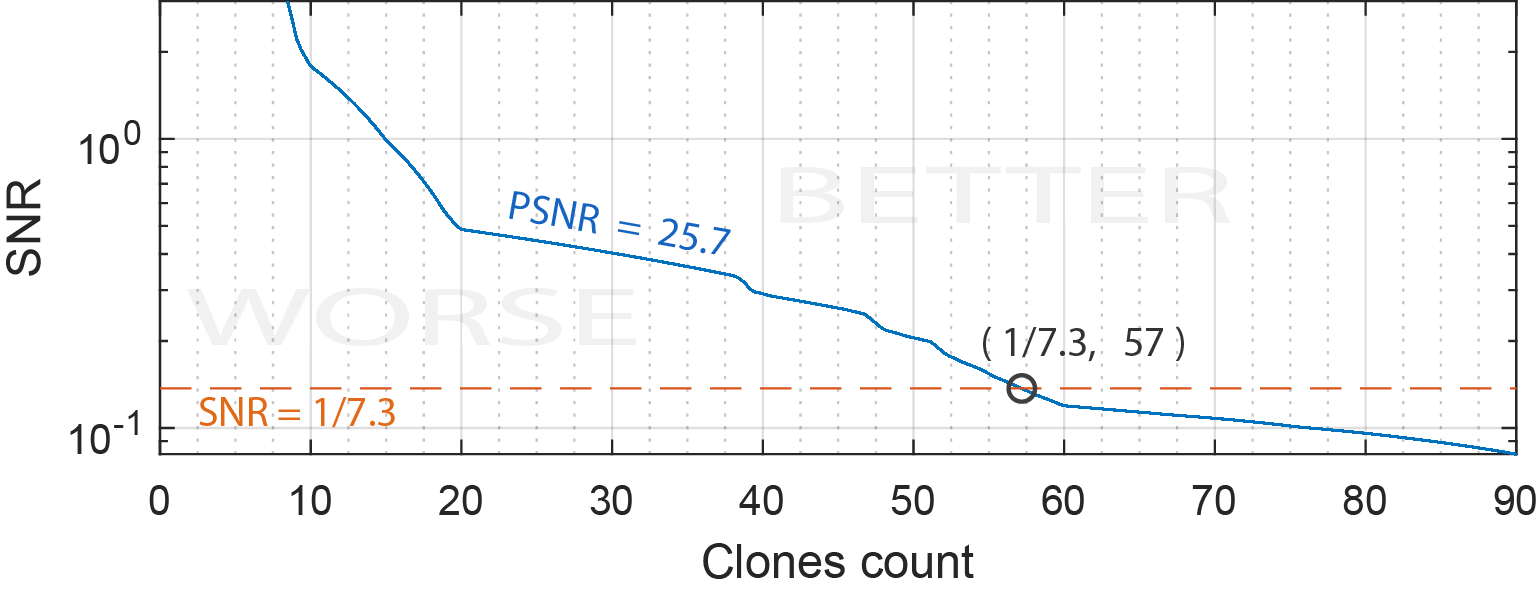}
	\caption{
	The dependency of minimal SNR on clones count required for the resolution enhancement in QT. The solid blue line indicates the performance of CT. Thus the areas above and below the blue line correspond respectively to \texttt{BETTER} and \texttt{WORSE} performence of QT compared with CT imaging (Fig.~\ref{fig:obscuredPupil}e, where PSNR = 25.7).
	}
	\label{fig:betterWorse}
\end{figure}

\begin{figure}[!ht]
	\centering
	\includegraphics[width=0.97\linewidth, trim=0.4cm 0.5cm 0.2cm 0cm]{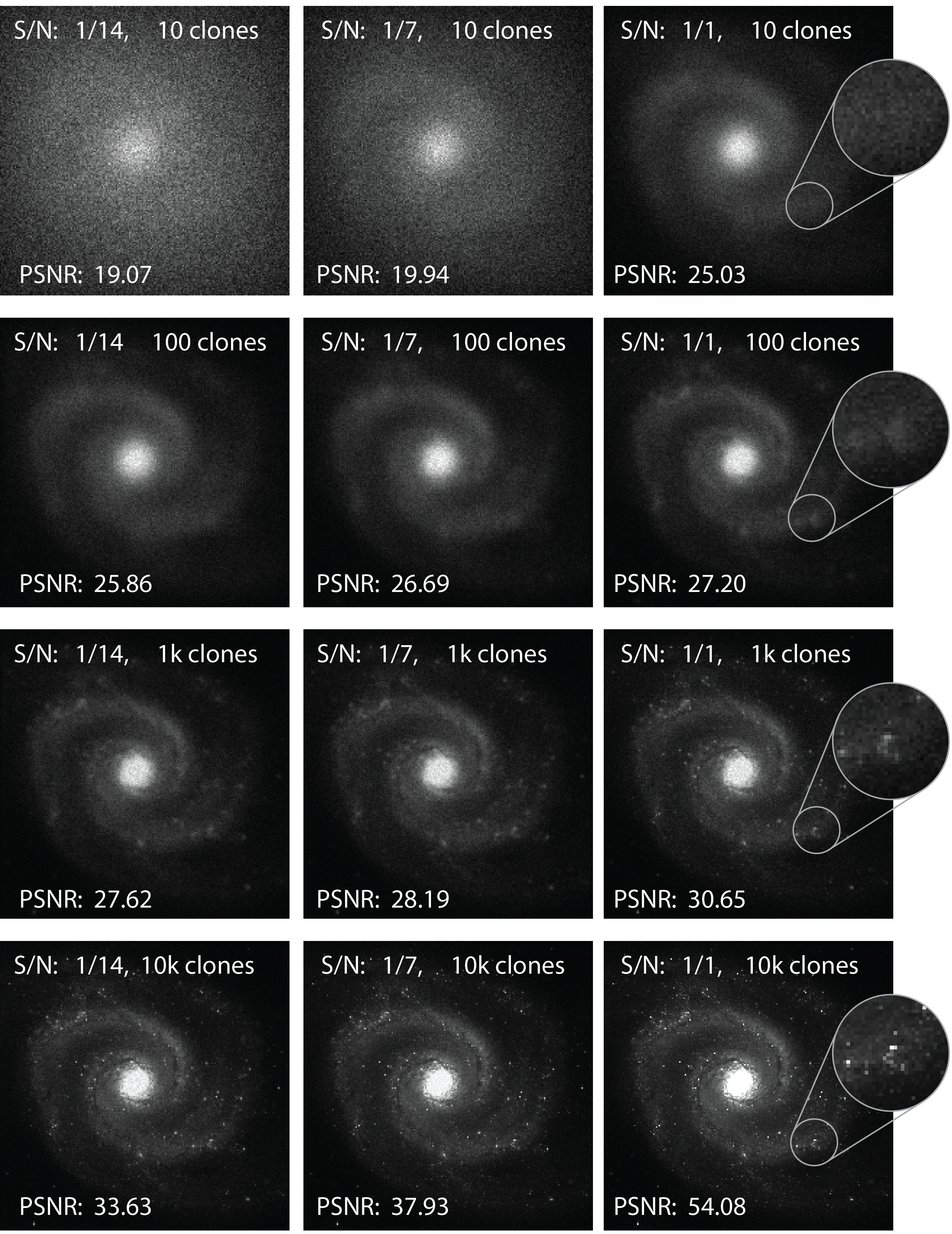}   
	\caption{
	Gain in the image quality as a function of clone count and SNR. Efficiency measure PSNR is depicted. Each row represent different clones count and each column represents different SNR. Zoom view shows how bright distinctive sources are restored.
	}
	\label{fig:outputImages}
\end{figure}

\noindent
where $I_{QT}(i,j)$ and $I_{red}(i,j)$ denotes the intensity of pixel at ($i,j$) in respectively QT simulated image and reference high-fidelity image. The PSNR and the mean centroid errors are depicted in Fig.~\ref{fig:surfs}. The place where both dependencies meet each other was retrieved and presented in Fig.~\ref{fig:betterWorse}. It shows the minimal requirements for the clones number and the SNR level, which should be satisfied to achieve the resolution enhancement in QT. This curve can be treated as a first guidance for selection of QT parameters. As seen from Figs.~\ref{fig:surfs} and \ref{fig:betterWorse}, assuming SNR $\simeq$ 7.3, in average $\gtrapprox$57 clones per 1 detected photon are necessary to produce an image sharper than CT. For $\sim$10k clones the image is virtually ideally restored, even for the lowest values of SNRs. The noticeable improvement of the outcome with higher clones count is related to the generally better estimation of the centroid when using the MF approach. For very small number of clones, the photons distribution is dominated by isolated photon-detection events and the MF output is strongly dependent on actual arrangement of registered clones. In contrast, for higher numbers of clones, the Gaussian becomes more uniform and therefore, the MF provides much more reliable estimations even if the SNR remains the same. In Fig.~\ref{fig:outputImages} we show exemplary resulting QT images for various SNRs and clones counts. The improvement in the image quality is easily noticeable as both parameters increase.

\vspace{2mm}
\section{Quantum Telescopes: technological feasibility}
\label{sec:technology}
\noindent
Below we summarize and discuss the feasibility of the QTs as emerging from the most recent literature. A constant progress in the QND increases the QT feasibility. There remains a possibility that some other kind of discriminator of spontaneous emission will be found. Possibly QND does not require to hold the photon in a resonant cavity for several hundreds microseconds, as it was claimed before~\cite{QNDnature}. According to~\cite{QNDnoCavity}, in the case of the cavity-free QND the photons momentum change is also negligible. However, this setup is suspected to produce false detections (<5\%) and cannot be miniaturized. Authors are presently working on a version easier to miniaturize (Keyu Xia priv. com.).

\begin{figure}[h]
	\centering
	\includegraphics[width=0.4\linewidth, trim=0 0.7cm 0 0] {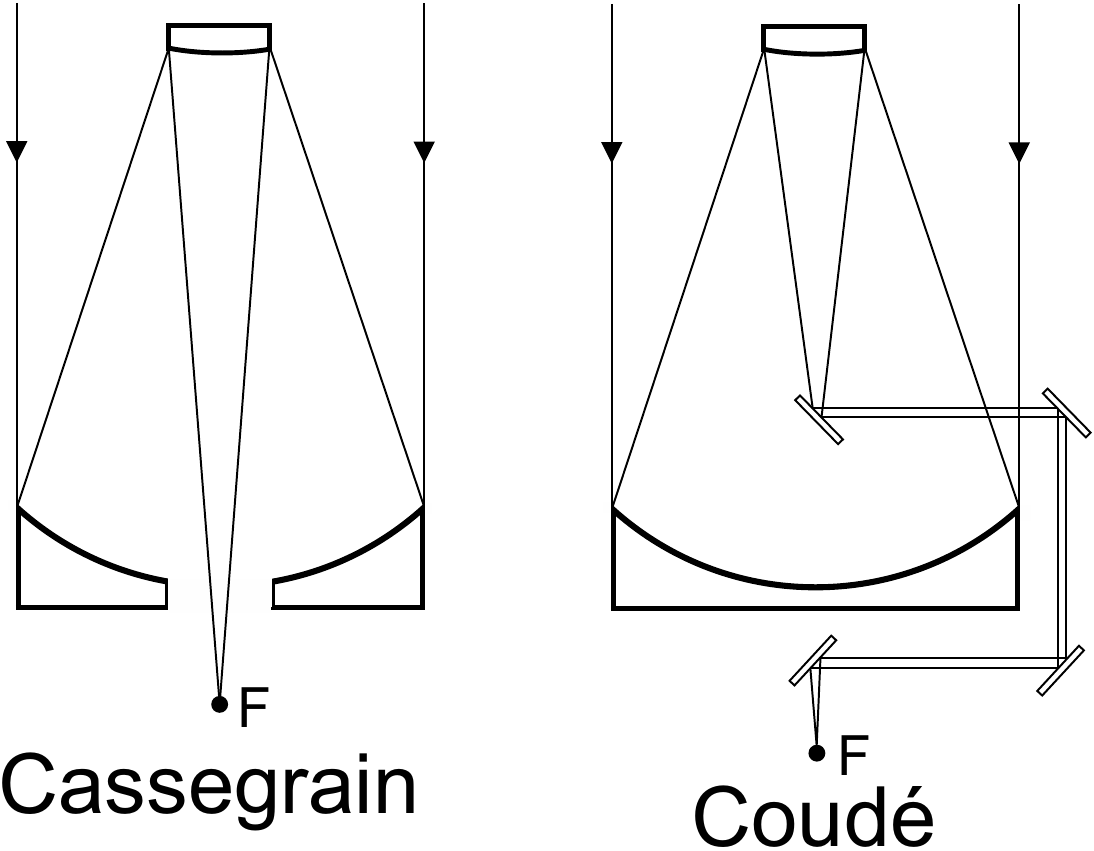}   
	\caption{
	\texttt{Cassegrain}\  and \ \texttt{Coudé}\  \ focus positions of the optical instrument (e.g. telescope). \texttt{Coudé} focus does not move as the telescope is reoriented.
	 }
	\label{fig:focus}
\end{figure}

\noindent
We see no strong limitations for the size of the device if it could be placed e.g. at the   \ \texttt{Cassegrain}  \ or\  \ \texttt{Coudé}\ \ plane optically conjugated to the pupil plane of the optical system near the focus~(Fig.~\ref{fig:focus}).

\noindent
Another technological issue concerning the QT is the spontaneous emission from the process of cloning (for details see~\cite{QT1}). As we showed, in our model, for a sufficiently large number of clones this problem could be overcome. However, in the general case of the QT some more sophisticated methods might be needed (e.g. different geometrical setup or filters).

\noindent
Given the recent progress in cloning, it might also become possible to clone photons from a wider wavelength span and at a greater amount~\cite{DeMartini2015}.

\vspace{6mm}
\section{Conclusions}
\label{sec:conclusions}
\noindent
In conclusion, the general idea of the Quantum Telescopes seems feasible and in the near future it might be possible to construct a technology demonstrator. In this paper we presented the updated schematic “toy model” of a QT system and the first quantitative results of simulations of such an imager, aiming at predicting the conditions which would guarantee its optimal performance. We found that it is generally more important to provide more clones than to reduce the noise background. Given the predicted SNR = 7.3, a satisfactory results are obtained from less than 60 clones. Using 10k clones, the signal is almost perfectly restored for any noise level. We encourage the interest and discussion on the idea of the QT, since even a small increase in the resolution might lead to a major breakthrough in astronomical imaging.

\vspace{6 mm}

\noindent
\textbf{Funding.} \ \ Marian Smoluchowski Research Consortium Matter Energy Future from KNOW. Polish National Science Center grant number UMO-2012/07/B/ST9/04425. Polish National Science Center, grant no. UMO-2013/11/N/ST6/03051. POIG.02.03.01-24-099/13 grant: GeCONiI - Upper Silesian Center for Computational Science and Engineering.

\noindent
\textbf{Acknowledgment.} \ \ The authors thank our colleague, Dr. W. Waniak for numerous useful discussions.

\end{document}